# A Sociotechnical Framework For Addressing Stigma and Designing Personalized Digital Health Products[1]

**Danielly de Paula**
Hasso Plattner Institute
danielly.depaula@hpi.de

**Daniel Jühling**
Hasso Plattner Institute

**Falk Uebernickel**
Hasso Plattner Institute

## Abstract

*Stigma, a recognized global barrier to effective disease management, impacts social interactions, resource access, and psychological well-being. In this study, we developed a patient-centered framework for deriving design requirements and interventions for health conditions subject to social stigma. This study introduces a patient-centered framework, grounded in sociotechnical systems theory, to create tailored interventions and design requirements for health conditions influenced by social stigma. We tested this framework through a mixed-method study on chronic pelvic pain patients. Our approach led to the identification of ten design requirements that encompass behavioral and psychological support and strategies for day-to-day living. The findings reveal a preference among CPP patients for priming and social support interventions. This study underscores the value of a systems-based perspective in healthcare, advocating for a nuanced, patient-centered approach that addresses the complex nature of health conditions affected by social stigma. It contributes to the ongoing discourse on integrating STS theory into healthcare frameworks, highlighting the need for targeted strategies to combat the complexities of stigma in patient care.*

## Introduction

Stigma is globally recognized as a significant obstacle to seeking healthcare (Scott et al. 2015), engaging in treatment (Rewerska-Juśko and Rejdak 2022), and adhering to prescribed therapies (Mulawa et al. 2021) across various health conditions (Stangl et al. 2019). Health-related stigma characterized by society's exclusion of individuals with certain health conditions severely impacts public health outcomes. For instance, it aggravates poor health by negatively affecting social interactions, access to resources, psychological and behavioral reactions (Hatzenbuehler et al. 2013). Despite the enormous impact of social stigmatization on public health, little is known about how to design digital health products for health conditions that are affected by stigma (Hofstraat and Brakel 2016).

The widespread use of mobile health (mHealth) presents a significant opportunity to address and potentially reduce health-related stigma. For instance, mHealth offers access to human support networks and Artificial Intelligence (AI) support (e.g., AI avatars) (Lee et al. 2022) as prior literature has revealed that AI systems can alleviate patients' perception of social stigma due to the feeling of having a judgment-free conversation (Pickard et al. 2016), whereas humans rely on interpersonal relationships for emotional support (Yan and Xu 2021). Additionally, mHealth offers advantages in addressing complex structural and psychosocial barriers by providing a platform to tailor interventions based on evolving user needs and preferences (Mulawa et al. 2021). Personalized interventions are particularly important in the context of health-related stigma due to the belief that they can reduce the barriers created by stigma, such as fostering patient empowerment through social support (Weitkamp et al. 2021) and reframing health conditions with the help of structured tools. However, health intervention design for stigmatized health conditions has been challenging for two main reasons.

---

[1]Preprint submitted to the International Journal of Human-Computer Interaction

First, there is the critical challenge of communicating with stigmatized patients. Patients are reluctant to share personal health data due to the fear of being judged or discriminated against (Graugaard 2017; Toye et al. 2014). The delicate nature of such health conditions requires a unique approach to communication, creating significant challenges for healthcare providers and mobile app manufacturers. This gap in understanding limits researchers' ability to extract correct requirements and inform the multi-level interventions required to meaningfully influence the stigmatization process.

Second, there is a lack of a process for designing interventions for health conditions that are affected by social stigma (Card 2022). Traditional patient-centered approaches focus broadly on tailoring healthcare to individual needs, preferences, and values (Gerken et al. 2022; Paula et al. 2022). However, when it comes to designing interventions for health conditions that are affected by social stigma, it is imperative to identify and emphasize those aspects that directly impact the experiences of stigmatized individuals. This includes understanding the unique psychological impacts of stigma, the social dynamics at play, and the barriers to healthcare access and adherence. Identifying these key elements will enable the development of more effective interventions that not only address the medical needs of patients but also the social and psychological challenges they face. This deeper, more focused approach to patient-centered design is essential for making significant strides in reducing the impact of social stigma in healthcare.

Based on the mentioned shortcomings, we recognize the need for a patient-centric approach with a focus on designing interventions for stigma-affected health conditions. Although the concept of patient-centered design has been integral to healthcare for decades (Staszak et al. 2021), its application in addressing social stigma requires a more nuanced understanding of the unique psychological impact of sigma. Accordingly, we propose the following research question:

**RQ:** What are elements of a patient-centered framework for deriving design requirements and interventions for health conditions subject to social stigma?

From a methodology perspective, we theoretically derived the framework based on social-technical systems (STS) theory and later tested it with patients who were diagnosed with chronic pelvic pain (CPP). CPP is a stigmatized health condition that affects up to 26% of females of reproductive age and 15% of all females worldwide (Lamvu et al. 2021). By integrating STS theory within a mixed-method design approach, our framework identified ten design requirements and the preferred intervention types of the patients. Our study contributes to the literature by proposing a patient-centered framework for stigmatized health conditions that is rooted in STS theory and operationalized through a mixed-method design approach, setting a precedent for future healthcare models.

## Literature Background

### *Navigating Stigma in Healthcare Delivery*

Over the years, the interpretation of "stigmatization" has changed significantly. Its historical roots trace back to ancient times, originating from the Greek word "stigma," which referred to physical marks branded on those deemed inferior in society, such as slaves or criminals (Goffman 1963). The modern interpretation of health-related stigma is defined as a 'social process, experienced or anticipated, characterized by exclusion, rejection, blame or devaluation that results from experience, perception or reasonable anticipation of an adverse social judgment about a person or group' (Weiss et al. 2006). Historically, social science researchers viewed stigma primarily as a byproduct of social interactions. Recent perspectives, however, have acknowledged the importance of self-perceived stigma as a significant source of distress. The Health Stigma and Discrimination Framework categorizes health-related stigma into enacted, anticipated, and internalized stigma (Stangl et al. 2019). Enacted stigma refers to actual discriminatory attitudes or social exclusion by the community towards a stigmatized individual. Anticipated stigma involves the individual's expectation of facing stigma, while internalized stigma deals with the stigmatized individual's absorption of negative stereotypes, often resulting in shame, guilt, and social withdrawal.

Stigma is particularly associated with chronic conditions as they often reshape the identity of the affected individuals (Weiss et al. 2006). For example, those with chronic mental health disorders might confront stigma related to perceived lifestyle choices or assumed personal responsibility for their condition, influencing their social identity and disease management. The impacts of stigma manifest daily, leading to social and psychological burdens, social isolation, diminished quality of life, and mental health issues (Hofstraat and Brakel 2016). Additionally, stigma affects broader aspects of disease management, such as health-seeking behavior, treatment adherence, and healthcare delivery, often exacerbated by policy issues that result in insufficient funding and support.

The body of literature refers to mHealth interventions as a promising approach to mitigating health-related stigma. For instance, by leveraging the anonymity and convenience of mobile platforms, mHealth creates a secure, private avenue for individuals to access support and information without fear of judgment, thereby countering anticipated stigma (Mulawa et al. 2021). Moreover, public-targeted interventions focus on dispelling misinformation and fears about stigmatized conditions, reducing enacted stigma (Liu et al. 2020). Additionally, these platforms enable users to find resources, manage their conditions, and join supportive communities, helping normalize various health conditions, thereby reducing internalized stigma (Mahar et al. 2022). Consequently, understanding how to design mHealth interventions for health conditions that are affected by stigma is a critical concern for researchers. Current challenges include fostering an environment where stigmatized patients are comfortable sharing their health data (Graugaard 2017) and the absence of a specifically defined process for designing such interventions (Card 2022). In our study, we argue for an enhanced approach to patient-centered care, viewing the issue through the lens of sociotechnical system theory. In the next section, we discuss an adaptation of the biopsychosocial model (BPSM), which is one of the most influential models for patient-centered care, based on a system-based theoretical lens.

## *A System-Based View to Enhancing Patient-Centered Healthcare*

Our study's foundation lies in sociotechnical system theory (Trist 1981), providing a lens to view the complex nature of healthcare and public health. This theory, initially developed to address industrial challenges in the British coal sector, emphasized the importance of self-managing work groups (Pasmore and Khalsa 1993). Its implementation brought significant improvements in productivity and worker satisfaction. Over time, the theory evolved to include participatory design principles, shifting from viewing workers as mere resources to recognizing them as individuals within a social context (Lyytinen and Newman 2008). Sociotechnical systems theory, fundamentally based on the integration of social and technical elements for successful outcomes, has since been applied in various research fields (Hofmann et al. 2023).

Within the healthcare field, sociotechnical systems theory has been used to enhance the biopsychosocial model by addressing its limitations (Card 2022). The biopsychosocial model, which is among the most influential frameworks for person-centered care, presented the determinants of health as a simple, nested system (Engel 1977). It gained widespread acceptance for its comprehensive view of health determinants, encompassing biological, illness, psychological, and sociological aspects (Roberts 2023). Criticisms of this model include its rigidity and lack of emphasis on technology in healthcare (Bolton and Gillett 2019). The integration of sociotechnical systems theory with the biopsychosocial model leads to the Biopsychosociotechnical model (BPSTM), acknowledging the constant interaction between social, technical, illness, and biological factors in health determinants (Card 2022).

The Biopsychosociotechnical Model serves as both a practical framework and a mid-range theory (Gregor 2006). It aims to assess and improve health determinants through participatory design and system-focused strategies. The BPSTM views health determinants as interconnected elements within a broader "system of systems," considering the interplay between social, technological, and biological factors. It suggests health conditions result from complex interactions between these elements. Therefore, enhancing the biopsychosociotechnical context involves optimizing all these dimensions to achieve a balanced state that improves health outcomes.

Overall, the Biopsychosociotechnical Model encompasses five key dimensions: Biological, Psychological, Disease/Illness, Social, and Technical, along with 28 health determinants. Table 1 compiles these health determinants, synthesizing relevant literature (Card 2022; Roberts 2023). These elements collectively outline critical areas for research and guide the development of health improvement strategies. In our research, we propose that the BPSTM provides a valuable framework for identifying essential design requirements for mHealth interventions for stigma-affected health conditions. Accordingly, we developed a patient-centered framework based on the BPTM for overcoming social stigma in healthcare (see Figure 1.

| Factor Dimension | Determinants of Health | Description |
| --- | --- | --- |
| **Biological** | Genetics | Conditions that can be traced back to genetic factors. |
| | Pathophysiological Changes | Pathophysiological processes that contribute to the disease. |
| | Physical Environment | Patient's environment (e.g., air quality). |
| | Nutrition, Sleep, Exercise | Dietary choices, sleep patterns, and physical activity. |
| | Medication | Pharmaceuticals for the management of the condition. |
| **Psychological** | Emotional States | Individual's emotional circumstances (e.g., stress). |
| | Expectations | Hopes and fears concerning disease management. |
| | Attitudes and Beliefs | Values and convictions that influence decision-making processes. |
| | Coping Strategies | Methods to manage and adapt to various forms of situations. |
| **Disease/Illness** | Pain | Perception of physical discomfort or distress in the form of pain. |
| | Disability | Impairments affecting physical, mental, or social functioning. |
| | Symptoms | Physical sensations besides pain that differ from one's usual health. |
| **Social** | Family, School, Work | The influence of family, educational environments, and occupation. |
| | Peer Relationships | The impact of interactions with peers and social networks. |
| | Context | The broader context of the present and past of the individual. |
| | Economic Forces | Economic factors that can either enable or hinder access to healthcare. |
| | Laws and Culture | Impact of legal regulations and cultural aspects. |
| | Norms | Societal norms, standards, and expectations towards an individual. |
| | Standard of Practice | Established guidelines for maintaining quality standards in healthcare. |
| | Health and Social Care Funding | Availability of necessary resources for the delivery of care. |
| **Technical** | Computer Hardware, Software and Networks | Utilization of electronics, software applications, and networks. |
| | Built Environment | Human-constructed physical environment (e.g., urban planning). |
| | Assistive Technology | Technologies designed to support individuals' quality of life. |
| | Furniture | Ergonomic design and arrangement of furniture for patient care. |
| | Forms | Documentation, paperwork, and record-keeping in healthcare. |
| | Pill Bottles | Design and usability of medication containers. |
| | Modes of Transportation | Transportation and its influence on access to healthcare. |
| | Medical Equipment | Medical devices used in diagnosis, treatment, and rehabilitation. |

**Table 1. Description of the determinants of health along their dimensions.**

## *The Case of Chronic Pelvic Pain*

Chronic Pelvic Pain as defined by the Royal College of Obstetrics & Gynaecology (RCOG), is a lower abdominal pain lasting over six months, not exclusively related to menstruation, intercourse, or pregnancy. This umbrella term encompasses various diagnoses such as Interstitial Cystitis (IC), Bladder Pain Syndrome (BPS), Vulvodynia (VD), and Endometriosis (EM). Individuals with these conditions report symptoms affecting urinary, bowel, myofascial, sexual, and gynecologic functions, significantly impairing their cognitive, behavioral, sexual, and emotional well-being. Pain symptoms in CPP vary widely among patients, ranging from sharp, searing pains to dull aches or cramping sensations, with some experiencing pain intensification during sexual activities or menstrual cycles (Lamvu et al. 2021).

Recent advancements in mHealth interventions for CPP patients have seen a growing interest in incorporating various psychotherapy techniques, as highlighted in a recent review paper (Mahar et al. 2022). These range from empathetic online assistants (Zippan et al. 2020) to anonymous question-asking forums managed by psychologists (Brotto et al. 2017) and peer support in online communities (Weitkamp et al. 2021). Mindfulness-based elements (Hess Engström et al. 2022; Weitkamp et al. 2021) and acceptance and com-mitment therapy (ACT) (Jennings and Apsche 2014) are also integrated, alongside psychoeducational components. Many interventions utilize mobile apps, supplemented by reminder calls or emails, and some include group video sessions with pelvic floor therapists (Jackowich et al. 2021) or involve reading material (Chisari et al. 2022). These approaches have shown improvements in well-being, anxiety, sexual function, and genital self-image (Weitkamp et al. 2021; Zarski et al. 2021).

However, CPP is a health condition that carries a significant social stigma, often due to its intimate nature and the societal discomfort surrounding discussions of pelvic health (Kapoor and McKinnon 2021). This stigma can lead to patients feeling embarrassed or ashamed, hindering communication about their symptoms and needs. Evidence shows that discussions between healthcare providers and their patients relating to sexuality are generally avoided (Graugaard 2017). Consequently, shared decision-making, crucial for optimal health outcomes, is hindered, leaving patients feeling isolated in their suffering. This limits patients' understanding of the nature of the problem, the treatment options, and the consequences. Furthermore, many CPP patients face invalidation from their social circles (Toye et al. 2014), exacerbating the condition. This creates significant challenges in designing effective health interventions. The economic impact of CPP is also noteworthy, with estimated costs in the US reaching $2.8 billion annually, and broader expenses exceeding $289 billion (Lamvu et al. 2021), emphasizing the need for efficient management and understanding in devising health interventions for CPP. In our study, we tested our patient-centered framework in a mixed-method study with CPP patients. The framework is presented in the next section.

## Research Design

Figure 1 presents our patient-centered framework theoretically rooted in the five dimensions of the Biopsychosociotechnical Model. The framework follows offers a system-based perspective of healthcare research in the context of social stigma. The five dimensions are not isolated; rather, they are interconnected and in constant interaction with each other, as evidenced by the connecting arrows. Our framework is operationalized through a mixed-method research design - i.e., the integration of both qualitative and quantitative research methodologies (Venkatesh et al. 2016). The framework allows for various mixed-method approaches (e.g., confirmatory, explanatory) tailored to the specific needs of the research (e.g., objectives, research questions, epistemological assumptions). For an in-depth analysis of the different strategies for mixed methods research, we refer to the comprehensive discussions in existing literature (Venkatesh et al. 2016).

The qualitative strand within the mixed-method approach aims to gain in-depth insights through an initial *engagement with patients*, which requires a foundation of trust and patient empowerment. Empowerment can be fostered in numerous ways. In our case, we collaborated with the Hale Community—a group dedicated to patient empowerment, consisting of over 10,000 members who feel confident to speak about their condition [2]. The quantitative strand aims to examine extensive datasets and *understand communities* of patients, employing data science techniques (Aragon et al. 2022). Here, using text mining techniques for analyzing online communities is recommended, as these communities offer the opportunity for patients to write their stories anonymously. The theoretical underpinnings—*knowledge base*—of the study are essential, serving as a lens for guiding research strategies.

To test the framework in the context of our study, we followed a concurrent mixed-method design, which is an approach that consists of two relatively independent strands that happen in parallel: one with qualitative questions and data collection and analysis techniques and the other with quantitative questions and data collection and analysis techniques (Teddlie and Tashakkori 2006). For this study, the two strands were (in dotted line): Repertory Grid (qualitative) and Computational Grounded Theory (CGT) (quantitative), both of which will be elaborated upon in subsequent sections.

Concerning the ethical aspects of this research, all participants were required to sign a consent form to participate in the RepertoryGrid study, whereas ethical approval was waived for the study described in the computational grounded theory section. The approval was waived because we used publicly available data and did not use any identifiable information. Despite ethical approval being waived, it was necessary to obtain written authorization from the platform's owner for the utilization of their data.

---

[2]https://www.halecommunity.com/

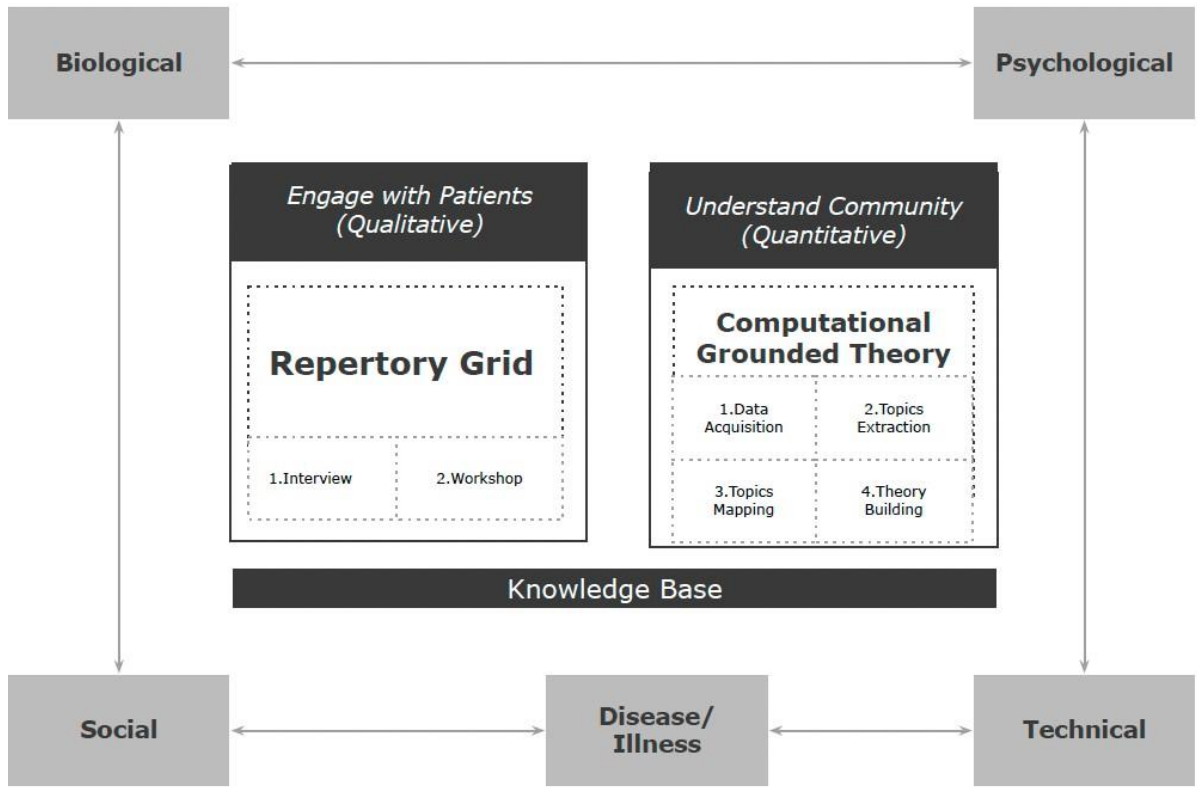

**Figure 1. A Patient-Centered Framework for Overcoming Social Stigma in Healthcare**

*Repertory Grid*

The Repertory Grid Technique is a cognitive mapping approach based on the personal construct theory (Kelly 1955), which is a constructivist theory that explains how people organize their environments in terms of cognitive personal constructs that are bipolar in nature, such as helpful - hurtful, emotional-analytical (Curtis et al. 2008). Since the personal constructs are elicited directly from the participants, the technique minimizes researcher bias (Tan and Gordon Hunter 2002), which encouraged many researchers to use the technique in various applications (Hille et al. 2023). In our study, we followed two major stages: interviews, and workshop - i.e., elements elicitation, construct elicitation, laddering, and linking elements and construct (Curtis et al. 2008). All participants signed a consent form before participating in the study. The RepGrid workshop aimed to identify relevant interventions for a digital health solution that aims to reduce pain interference to the social and mental well-being of patients with CPP.

**Pre-workshop interviews**

The interviews were designed to uncover patients' preferences for various interventions. We conducted 10 semi-structured interviews with patients previously diagnosed with chronic pelvic pain, using video calls in May and June 2023 for this purpose. While most interviews were in English, one was conducted in Spanish, languages in which the interviewer is fluent. To facilitate our coding and data analysis, we utilized Atlas.ti software. Adhering to established qualitative research methodologies, such as constant comparison and iterative conceptualization, we extracted meaningful insights (Gioia et al. 2013; Urquhart et al. 2010). For instance, this approach led us to determine that priming for emotional and utilitarian support, along with social support (peer-to-peer and expert support), were the most favored interventions among patients. Given its significant relevance for creating community-based digital health solutions, we decided to focus further on social support for our next phase.

**RepGrid Workshop**

The workshop's goal was to identify and prioritize social support interventions based on peer-to-peer and expert support interventions. A total of 10 new patients attended a two-hour online session. During this, two RepGrids were developed, and their construction was documented on a Miro board. The initial step involved generating elements for the RepGrid focused on peer-to-peer support, where participants added examples such as buddy programs and mentorship. For construct elicitation, we employed the triadic method, widely regarded as more effective than the dyadic approach for generating useful constructs (Curtis et al. 2008). This involved participants selecting three elements and distinguishing the third from the first two in terms of characteristics, such as emotional versus rational intervention. In the final stage, involving "laddering and linking elements and constructs," participants were prompted to explain their reasoning behind each construct and categorize them as either "less desirable" or "desirable" on the RepGrid. They then rated each element on a scale from 1 (less desirable) to 5 (desirable) according to the elicited constructs. This process was replicated for identifying and ranking expert support interventions, such as online webinars with healthcare professionals.

## *Computational Grounded Theory*

The purpose of CGT is to produce theoretical insights from patterns identified using computational techniques (Berente et al. 2019). It offers a powerful tool for analyzing large datasets and uncovering nuanced insights by combining traditional grounded theory methodology with advanced computational techniques (Shaila Miranda, Nicholas Berente, Stefan Seidel, Hani Safadi, Andrew Burton-Jones 2022). It is a well-established approach for theorizing from large datasets. For this study, we followed the recommendations from scholars to operationalize CGT (Berente et al. 2019; Ojo and Rizun 2021). Their approach involved the following activities: data acquisition, extracting topics, lexicon framing, and theory construction.

**Data Acquisition & Preparation**

Our dataset was sourced from an Italian forum dedicated to cystitis patients [3]. This forum is managed by Cistite.info APS, an organization committed to aiding individuals with urogenital conditions and promoting health education, particularly focusing on women's well-being. The forum's content is publicly accessible, allowing anyone to read posts without needing an account. All posts are in Italian language. To obtain permission for the analysis of the forum's content, we contacted the forum administrators, who gave us their explicit consent. To align with our research objectives, we asked for an expert in CPP to help us select relevant subforums. These subforums covered a range of topics, including symptoms, hygiene, travel challenges with urogenital pathologies, sexual and relationship issues, and specific diagnoses and treatments. For our analysis, we compiled posts only from these specifically relevant subforums. As of March 21, 2023, the forum had a total of 343,853 posts across 4,839 threads, contributed by 10,462 registered members. The subforums selected for our study contained 75,011 posts.

In our study, we employed the Scrapy Python package [4] for data collection. Scrapy is a sophisticated and flexible framework specifically designed for web crawling and scraping, enabling the efficient extraction of structured data from websites. This framework allows for the creation of "spiders," which are scripts programmed to navigate specific websites and systematically harvest relevant information. For our research on the Cistite.info forum, we developed two distinct spiders. The first spider was tasked with gathering hyperlinks to all threads across the forum's subforums. Following this initial step, a second spider was deployed to traverse these identified threads, scraping all the posts within them. The data extracted from each post included the text of the post, its publication date, the specific thread, and the corresponding subforum to which the thread belonged. Importantly, to maintain the privacy of forum participants, usernames were deliberately not collected, thereby ensuring the anonymity of the post authors in our analysis.

In the initial phase of data pre-processing, we focused on refining the raw HTML content of the forum posts. This involved the removal of all quoted sections, which were identified by the HTML "<blockquote>" tag.

---

[3] https://cistite.info/forum/index.php
[4] https://scrapy.org/

Utilizing regular expressions, we extracted and discarded these tags and their enclosed content, ensuring that only the original text authored by the post's creator remained. Subsequently, the posts received further cleansing to eliminate any residual HTML tags. Additionally, we standardized the text by converting it to lowercase and removing double quotes, thereby simplifying string representations and avoiding potential parsing issues. Given the Italian origin of the posts, we employed Google's Translation API for language translation into English. To accommodate the API's character limit of 4,999 characters per submission, longer posts were segmented into smaller, coherent text sections. These divisions were carefully executed to begin at the start of new sentences, maintaining the semantic integrity of each segment. To focus our analysis on substantial content, we imposed a minimum character threshold of 200 for each post. This criterion aimed to exclude brief, non-informative messages, such as expressions of thanks or praise, which do not contribute significantly to our research objectives. Following these pre-processing steps, a corpus of 43,655 posts was compiled for in-depth analysis. The translated content encompassed a total of 8,047,714 words, featuring 43,426 unique words.

**Extracting Latent Topics**

For the extraction of latent topics from our dataset, BERTopic was employed due to its superior selective approach in topic categorization, particularly in comparison with other topic modeling methods like Latent Dirichlet Allocation (LDA) (Grootendorst 2022). This selective approach significantly improves BERTopic's capability to identify and clearly define unique, non-overlapping topics. This enhanced precision in topic categorization was a decisive factor in choosing BERTopic for our research.

BERTopic uses five phases to produce topic distribution based on a set of documents. First, Document Embedding converts text into numerical representations using SBERT. Next, UMAP is employed in Dimensionality Reduction to manage high-dimensional data, aiding in forming clusters of similar documents (McInnes et al. 2018). Clustering is achieved through HDBSCAN, grouping data based on density and minimizing misassignments to enhance quality. In vectorization, documents within a cluster merge, analyzing word frequencies at the cluster level, normalized to cluster size. Lastly, Topic Representations are formed using class-based TF-IDF (Grootendorst 2022), evaluating word significance in clusters for more precise topic descriptions.

The BERTopic topic modeling process required some adjustments to accommodate the unique characteristics of our dataset. For instance, we addressed inaccuracies in Google's translation, like idiomatic expressions and informal language. The model was iteratively refined, identifying misinterpreted words and usernames, creating a list of 309 such terms. This list, combined with English stopwords, was used to filter out irrelevant terms in the vectorization process. To suit our large dataset of 46,355 posts, we increased the minimum topic size from 10 to 30 documents and adjusted UMAP's nearest neighbors from 15 to 20 for a better balance between global and local structures, leading to more evenly distributed topic sizes.

We evaluated our topic model using the Topic Coherence metric, specifically the Conditional Normalized Point-wise Mutual Information (CNPMI) from the Gensim Python package, known for reflecting human-like topic interpretations (Lau et al. 2014). Additionally, we used Topic Diversity (Dieng et al. 2020) to measure uniqueness among the top 25 words in each topic, indicating how distinct the topics are. Combining these, we calculated Topic Quality, a metric assessing both coherence and diversity, to determine the ideal number of topics in BERTopic. Initially, BERTopic produced 53 topics, exceeding our target range for ease of human interpretation. To improve this, we gradually reduced the topic count, starting from 45 and decreasing by 5, recalculating Topic Quality at each step to find the optimal balance of coherence and diversity. After careful consideration of both diversity and coherence, we decided on a number of 25 topics, of which 23 were deemed valid and selected for further analysis. The values of Topic Coherence, Topic Diversity, and Topic Quality are detailed in Table 2.

| Number of Topics | Topic Coherence (TC) | Topic Diversity (TD) | Topic Quality (TQ) |
| --- | --- | --- | --- |
| 45 | 0.1373 | 0.8045 | 0.1105 |
| 40 | 0.1401 | 0.8436 | 0.1182 |
| 35 | 0.1355 | 0.8471 | 0.1148 |
| 30 | 0.1280 | 0.8517 | 0.1090 |
| 25 | 0.1283 | 0.8958 | 0.1150 |
| 20 | 0.1246 | 0.9000 | 0.1122 |
| 15 | 0.1465 | 0.9000 | 0.1318 |
| 10 | 0.1570 | 0.9000 | 0.1413 |
| 5 | 0.1981 | 0.9750 | 0.1932 |

**Table 2. Topic evaluation metrics for the BERTopic model results.**

**Topics Mapping & Theory Building**

The analysis of the 23 topics involved two key steps for interpretation and understanding. First, we examined the most relevant posts within each topic, with the support from an Italian-speaking expert in CPP to interpret colloquial language and nuances. Second, we aligned these topics with the BPST model. This involved two researchers independently categorizing topics (first-order coding) into one of the BPST model's five dimensions, based on the same set of representative forum posts. After an in-depth discussion, the researchers reached a consensus on the mapping of each topic. Topics that didn't neatly align with any specific dimension were set aside for further analysis. Table 4 shows the results of the mapping. This process allowed us to derive substantive theory though a primarily inductive process that resulted in the extension of an existing lexicon (Berente et al. 2019), and the identification of design requirements. To establish a solid foundation for these design requirements, the process of design requirement extraction was guided by previous research (Slater et al. 2017).

# Results

## *Mapping topics to the BPST model*

Table 4 illustrates the result of the process of mapping the topics with the BPST model, adhering to established scientific procedures for data visualization (Gioia et al. 2013). The second-order codes refer to the latent topics extracted from our dataset, which is indicated by the number within the parenthesis. Firs-order codes refer to the health determinants that belong to the BPSTM model. The codes highlighted in bold represent an extension of the model as a result of this study. The third column refers to the factor dimensions from the BPST model - some topics were mapped to more than one dimension.

The **Biological Factor Dimension** uniquely encompasses six topics linked to two health determinants. Firstly, the topics related to *Nutrition, Sleep, Exercise* involve discussions about digestive health (T5), the influence of physical activity on the condition (T8), and homemade treatment methods (T23). Secondly, the *Medication* determinant is addressed in topics discussing treatment options for viral/bacterial infections (T11) and the adverse effects of antibiotics in dental extractions (T19).

The **Disease/Illness Factor Dimension** covers four topics addressing two health determinants. Under the determinant of *Symptoms*, we found topics discussing various burning sensations (T7), side effects of medications (T13), and symptoms associated with HPV and other diseases (T15). Additionally, one topic pertains to the *Pain* determinant, focusing on stress-triggered comorbidities (T9).

The **Social Factor Dimension** encompasses three topics addressing two health determinants, with one being a novel inclusion. Under the determinant of *Standard of Practice*, topics included discussions about doctor recommendations and testimonials (T17), as well as narratives about patients' experiences with various treatments and healthcare providers, and significant life events (T22). Our analysis also suggests the introduction of *Partner Relationship* as a new health determinant, recognizing the significant challenges patients face in maintaining intimacy with romantic partners (T1). This addition underscores the need for manufacturers and researchers to distinguish between peer and partner relationships, acknowledging the latter as a critical aspect of social health for patients diagnosed with CPP.

The **Technical Factor Dimension** includes two topics related to two health determinants, with one being a new addition. The first topic (T20) addresses the significance of ergonomic *Furniture* for rest and the supportive role of *Assistive Technology* in treatment. Furthermore, our study revealed the critical impact of *Clothing* as a health determinant. Many patients reported specific fabrics or types of clothing exacerbating burning sensations in the pelvic area (T2). This underscores the importance of tailored clothing advice for individuals with CPP, leading us to propose Clothing as an essential health determinant for CPP patients.

The **Psychological Factor Dimension** encompasses two topics associated with the determinant of *Coping Strategies*. Topic T14 highlighted the therapeutic value of pets, with users referring to them as "one of the most effective therapies to lift everyone's spirits." Furthermore, discussions about coping with vulvodynia/vestibulodynia/vestibulitis (T3) revealed significant insights. A user shared their experience, stating, "(...)I realized I had been living passively, with my vulvodynia dictating my days instead of me taking control. (...) While I'm not completely healed, I do feel less oppressed, less fragile, and less insecure due to vulvodynia." This highlights the importance of proactive coping strategies in managing the psychological aspects of conditions related to CPP.

At the juncture of the **Biological and Technical Factor Dimensions**, two topics emerged. The first, Topic T0, involves discussions about medications influencing vaginal pH and flora. This was categorized under *Medication* in the Biological dimension and *Medical Equipment* in the Technical dimension, reflecting conversations about self-measurement of vaginal pH. The second intersecting topic, Heat to Alleviate Contracture, Effects of Temperature on the Condition (T12) falls under *Physical Environment* within the Biological dimension, addressing the impact of cold weather on CPP. Additionally, this topic was classified under *Assistive Technology* in the Technical dimension due to discussions about using heat pads and patches, highlighting user experiences with these aids.

In the overlap between the **Biological and Social Factor Dimensions**, we identified three topics. First, Pregnancy and Contraceptive Measures (T4) was categorized under *Family, School, Work* within the Social dimension, as it involved discussions on family formation following childbirth. From the Biological perspective, a new code, *Physiological*, was created to encompass discussions about the physiological aspects of pregnancy, a theme that previously lacked a fitting code in the Biological Factor Dimension. The second topic bridging the Biological and Social dimensions was Vaginal Lubrication and Lubricant Products (T16). This topic was classified under *Medication* and *Pathophysiological Changes* in the Biological dimension due to discussions on countering decreased vaginal lubrication (due to the condition) with pharmaceuticals. Within the Social dimension, the topic was linked to the new *Partner Relationships* code, reflecting conversations about how changes in vaginal lubrication impact the patients' sexual life. Furthermore, discussions about Beauty standards, pubic hair removal, and its interaction with the condition (T21) show how shaving can exacerbate patients' discomfort as a result of a *Pathophysiological Change*. However, users also pointed out that shaving is a cultural norm "Imagine if they hadn't promoted the idea that a shaved woman is more beautiful than a natural one. We could have spared ourselves from countless societal pressures and expectations"; therefore, T21 was also mapped to Law & Culture within the Social dimension.

Within the overlap of the **Disease/Illness and Psychological Factor Dimensions**, we identified a topic addressing *symptoms* and *coping strategies* for chronic pelvic pain on a more general level. This topic, referred to as Potato (a colloquial Italian synonym for vulva), coping with the condition (T10) encompasses discussions about the general symptoms of chronic pelvic pain and various methods used to manage the condition.

The topic Dilators, Self-Massage, Reverse Kegels (T6) uniquely intersected the **Biological, Technical, and Psychological Factor Dimensions**, receiving one code from each. From the Biological dimension, it was coded under *Nutrition, Sleep, Exercise* due to discussions about specific exercises such as reverse Kegels. In the Technical dimension, it was classified under *Assistive Technology* and *Medical Equipment* because of conversations about tools such as dilators and massagers. Additionally, it was linked to the *Emotional States* code in the Psychological dimension, capturing the mental anguish associated with using these devices on painful body parts.

Aligning the BPST model with topics from our dataset enabled us to discern the most pertinent health determinants for CPP patients. Our findings reveal that out of 28 established health determinants, 15 were relevant to CPP, and 3 new determinants emerged, warranting their inclusion in the revised BPST model. Table 3 details the relevant health determinants for CPP patients, marked with an asterisk (*), and highlights the newly identified determinants (**) for integration into our adapted model, as determined by our analysis.

| Factor Dimension | Health Determinants |
| --- | --- |
| **Biological** | Genetics<br>Pathophysiologic Changes*<br>Physical Environment*<br>Nutrition, Sleep, Exercise*<br>Medication*<br>Physiological** |
| **Psychological** | Emotional States*<br>Expectations<br>Attitudes and Beliefs<br>Coping Strategies* |
| **Disease/Illness** | Pain*<br>Disability<br>Symptoms* |
| **Social** | Family, School, Work*<br>Peer Relationships*<br>Context<br>Economic Forces<br>Laws and Culture*<br>Norms<br>Standard of Practice*<br>Health and Social Care Funding<br>Partner Relationships** |
| **Technical** | Computer Hardware, Software and Networks<br>Built Environment<br>Assistive Technology*<br>Furniture*<br>Forms<br>Pill Bottles<br>Modes of Transportation<br>Medical Equipment*<br>Clothing** |

**Table 3. Biopsychosociotechnical Model Factor Dimensions and associated health determinants**

| Second-order Codes (Topic Number) | First-order Codes / Determinants of Health | Biopsychosociotechnical Model Factor Dimensions |
| --- | --- | --- |
| Vaginal pH and flora, different pharmaceuticals and their effect and side-effects (0) | Medication<br>Medical equipment | Biological<br>Technical |
| Coping with the condition, relationships and intimacy (1) | **Partner Relationships** | Social |
| Clothing fabrics and their influence on the condition, washing products (2) | **Clothing** | Technical |
| Coping with vulvodynia/vestibulodynia/vestibulitis (3) | Coping Strategies | Psychological |
| Pregnancy and contraceptive measures (4) | Family, School, Work<br>**Physiological** | Social<br>Biological |
| Constipation, bowel movements, hemorrhoids and their effect on the condition (5) | Nutrition, Sleep, Exercise<br>Medication | Biological |
| Dilators, self-massage, reverse kegels (6) | Emotional States<br>Nutrition, Sleep, Exercise<br>Assistive technology<br>Medical equipment | Psychological<br>Biological<br>Technical |
| Burning sensation as a symptom, heartburns (7) | Symptoms | Disease/Illness |
| Different sports and exercises and their effect on the condition (8) | Nutrition, Sleep, Exercise | Biological |
| Comorbidities: fibromyalgia and pudendal neuropathy (9) | Pain | Disease/Illness |
| Potato (a colloquial Italian synonym for vulva), coping with the condition (10) | Symptoms<br>Coping Strategies | Disease/Illness<br>Psychological |
| Viral/Bacterial Infection & Immune System (11) | Medication | Biological |
| Heat to alleviate contracture, effects of temperature on the condition (12) | Physical Environment<br>Assistive Technology | Biological<br>Technical |
| Itching as a symptom or side-effect of Dobetin (13) | Symptoms | Disease/Illness |
| Pets as companions (14) | Coping Strategies | Psychological |
| HPV and other venereal diseases (15) | Symptoms | Disease/Illness |
| Vaginal lubrication and lubricant products (16) | **Partner Relationships**<br>Pathophysiological Changes<br>Medication | Social<br>Biological |
| Recommendations and testimonials about doctors (17) | Standard of practice | Social |
| Diets and the effect of Laroxyl on body weight (18) | Nutrition, Sleep , Exercise<br>Medication | Biological |
| Having teeth removed and how preventative antibiotics affect the condition (19) | Medication | Biological |
| Resting positions and their effect on the condition, cushions as treatment (20) | Furniture<br>Assistive Technology | Technical |
| Beauty standards, pubic hair removal and its interaction with the condition (21) | Pathophysiological Changes<br>Law & Culture | Biological<br>Social |
| A user's patient journey, trying different treatments and doctors, and life events (22) | **Standard of practice** | Social |
| Chili/Capsaicin's effect as a home-made treatment for women's urologic conditions (23) | Nutrition, Sleep, Exercise | Biological |

**Table 4. Results of the coding process starting from second-order codes/topics as produced by BERTopic. First-order codes newly devised in the process of the model expansion are printed in bold type.**

*Identifying design requirements*

The analysis of forum discussions among CPP patients has led to the identification of ten design requirements (DRq) explained in this section. Our findings emphasize the wide-ranging and interrelated needs of CPP patients, going beyond traditional medical interventions to include behavioral considerations, psychological support, and strategies for day-to-day living. Notably, our analysis also underscores the importance of environmental factors in health management, such as adapting to the ways in which colder weather can exacerbate CPP symptoms. This broad perspective on health highlights the multifaceted nature of CPP and the necessity for holistic management strategies.

Given the often insufficient seriousness with which healthcare professionals treat CPP patients, it is **crucial to facilitate access to a network of specialized healthcare practitioners (DRq1)**. This need is derived from discussions in T17 and T22, where users voiced an urgent need for medical professionals who not only have a comprehensive understanding of CPP but also approach patients with empathy and are knowledgeable about effective treatment options. It was mentioned several times that trusted professionals are rare, emphasizing the need for broader access to specialized care.

Another key need identified from our analysis involves **providing information about common symptoms and related diagnoses of chronic pelvic pain (DRq2)**. This requirement, underscored in T7, T10, and T21, reflects users' desire for a deeper understanding of CPP symptoms, such as burning sensations in the vulval area and discomfort related to pubic hair. There is a clear demand for insights into the progression of CPP symptoms and strategies for lifestyle adaptation to alleviate them. Beyond just recognizing symptoms, forum discussions in T7, T9, and T15 reveal a significant gap in understanding related diagnoses. Users frequently report varied symptoms and express uncertainty about their connection to CPP. They also show a keen interest in learning about common co-occurring conditions, such as depression, indicating a need for comprehensive information that bridges symptom recognition with broader diagnostic awareness. This highlights an overarching need for enhanced knowledge and understanding among CPP patients.

Another critical aspect emerged from our analysis: the need to **educate patients about pharmaceutical side effects and mitigation strategies (DRq3)**. This necessity, highlighted in T0, T11, T13, and T19, addresses the adverse reactions from medications often prescribed to CPP patients. However, it is evident from the forum discussions that patients are actively seeking ways to complement medication with alternative solutions. They expressed a keen interest in being **informed about self-care and homemade remedies (DRq4)**, underscoring a desire for more autonomy in managing their condition. Examples include self-massage techniques and heat pads for alleviating pain, as noted by users. This requirement, synthesized from T6, T8, T12, and T23, reflects a strong drive among CPP patients to take charge of their health and well-being, moving seamlessly from a passive approach to focusing on self-empowerment.

Following the identified needs for information and empowerment in managing CPP, our analysis also underscores the importance of **creating a secure communication platform for CPP patients (DRq5)**. This need, highlighted in T1 and T2, stems from the desire of patients for a judgment-free zone where they can share experiences, exchange advice, and find solace in the empathy and understanding of others. Given that CPP profoundly impacts sexual health, there is a critical demand for **practical, emotional, and interpersonal guidance on intimacy (DRq6)**, rooted in discussions from T1, T6, T16, and T21. This requirement calls for resources that provide clear and comprehensive advice, helping patients navigate the complexities of intimacy while considering the physical and emotional facets of their condition. Additionally, there's a need to **inform about the medical and social nuances of pregnancy in relation to CPP (DRq7)**. This highlights the need for tailored resources and support mechanisms addressing CPP patients' unique challenges and uncertainties during pregnancy. These three design requirements together emphasize the sensitive and personal nature of living with CPP, pointing towards a holistic approach in addressing the multifaceted impacts of the condition.

Acknowledging the substantial life changes experienced by those with CPP, it becomes essential to **guide patients on lifestyle adaptations and their impact on CPP (DRq8)**. Key areas of focus include understanding the relationship between diet, metabolic processes, the clinical progression of CPP (T5 and T18), and strategies to mitigate pain during cold weather (T12). Additionally, there's a significant interest in advice on tools and devices such as cushions or specialized furniture for comfortable sitting postures and optimal pelvic positioning (T20). Furthermore, the **impact of clothing and fabric choices on CPP (DRq9)** is a critical consideration, as highlighted in T2. Clothing choices for CPP patients extend beyond fashion, significantly affecting their symptoms. Providing comprehensive advice on suitable fabrics and clothing options is therefore crucial. Lastly, considering the mental health implications of CPP, there's a need to **offer psychological techniques and philosophical approaches for coping with the condition (DRq10)**, a need derived from insights in T3, T10, and T14. This holistic approach to managing CPP underscores the importance of addressing not just the physical but also the emotional and mental aspects of the condition.

*Identifying interventions*

The goal of this step was to identify relevant interventions for a community-based digital health solution that aims to reduce pain interference with the social and mental well-being of patients with CPP. To do that, we conducted a Repertory Grid workshop. First, to obtain in-depth knowledge about the patient's preferences for health interventions, we conducted pre-workshop interviews. As a result, treatment options, challenges, and needs options were identified.

The **treatment** options pursued by participants can be categorized into three main groups: physical therapies, lifestyle changes, and medication use. Physical therapies were a common approach, with participants engaging in activities such as massages, Pilates, and specific therapy exercises. This category also included the use of electrical stimulation and physiotherapy. Some participants expressed a preference for professional assistance in these therapies due to concerns about incorrectly performing them and potentially exacerbating their condition, though this often led to higher costs. Lifestyle changes formed another crucial category. These included walking in nature, doing yoga, journaling to organize weekly plans and appointments, and dietary changes. The third category involved medication and specialized treatments. Participants reported using localized medications and anesthetic creams, and also medication to treat associated conditions (e.g., migraine). Some explored novel methods such as electric pulsation therapy, highlighting a willingness to try diverse treatment modalities. T*hus, the predominant treatment strategies for the participants focused on therapeutic exercises, such as massages, and mood-regulating activities, such as nature walks and yoga.*

Individuals with CPP encounter a range of **challenges** that can be broadly categorized into three main areas: lifestyle changes, mental health impacts, and social and medical acceptance. First, integrating treatment into daily life presents significant hurdles. Many CPP patients lead busy lives, balancing academic or professional responsibilities with their health needs. Establishing a consistent routine for therapy exercises is a common struggle, often disrupted by fluctuating mental states. Periods of traveling are even more challenging, especially when ensuring access to necessary medications and maintaining privacy for daily exercises. Second, the mental health impact of CPP is profound. The constant pain and fatigue affect physical resilience and significantly impacts mental strain. Patients mentioned that having to explain their limitations to others repeatedly is emotionally draining and isolating. This aspect is often aggravated by a lack of understanding and empathy from others, including close acquaintances, leading to feelings of frustration and anger. Third, the challenges of social and medical understanding are evident in the interviewees' experiences of misdiagnosis and the necessity for self-advocacy. The lack of support from some family members forces individuals to take sole responsibility for advocating for their own health and well-being. Romantic relationships are also challenging for individuals with CPP, especially when dealing with intimate aspects such as penetrative sex. This requires sensitive handling and can frequently lead to further difficulties within the relationship. Many participants also emphasized the effort to obtain an accurate diagnosis. The interviews revealed to have received multiple misdiagnoses, with some individuals only receiving a correct diagnosis a decade after their initial symptoms appeared. *Thus, the main challenges faced by the participants were how to establish a consistent routine and how to manage the emotional impact of the conditions.*

Our findings also revealed the different **needs** of the patients based on their maturity in the management of the condition. Patients whose treatment routine is already established (e.g., daily exercises, visits to a psychologist) seek a digital health solution that will complement their current knowledge with tips about everyday life, such as suitable clothing choices or recommendations for doctors. Overall, these patients wish for supplementary support that enhances their existing routine and psychological care. On the other hand, patients without a set routine or who require additional support exhibit different needs. When a routine is not established, patients tend to need more regular supportive interactions, reminders for the exercises, and positive reinforcement strategies to cope with mental health disorders. Furthermore, these patients emphasize the need for tools that facilitate self-reflection and personal growth, suggesting a desire to understand and manage their condition more deeply. The most mentioned needs referred to having access to mood boosters, frequent interactions with healthcare professionals, and more personal communication with peers. *Thus, this indicates that the main needs of the participants evolve around peer-to-peer interactions and expert guidance.*

Further analysis of the findings revealed the **intervention preferences** of participants in a digital health solution. First, the findings suggest the importance of offering *priming interventions* focused on emotional support, such as mood boosters, and practical support, such as reminder systems for taking medication or doing exercises. This is particularly important for patients who are still in the process of establishing a routine. Furthermore, the analysis highlights the critical need for *social support interventions*, including peer-to-peer support mechanisms such as mentorship programs, and access to expert guidance through formats such as webinars. By integrating these diverse forms of support - emotional, utilitarian, peer, and expert - into a digital health solution, patients can receive a more holistic and comprehensive approach to managing their health, ultimately leading to better outcomes and enhanced quality of life. For the next step, the RepGrid workshop, we decided to focus on the most important type of intervention based on our analysis: social support.

We then conducted the RepGrid workshop to identify and rank two different types of social support that were identified during the pre-workshop interviews - peer-to-peer support and practitioner support. In total, 6 constructs and their bipolar counterpart (e.g., rational-emotional), 9 peer-to-peer support, and 9 practitioner support were identified. Table 5 illustrates the list of social support interventions ranked by means of importance.

| **Rank** | **Peer-to-peer support** | **Practitioner Support** |
| --- | --- | --- |
| **1** | Mentorship | Online webinar + QA |
| **2** | A platform for building connections with other patients | Guidelines to follow at home |
| **3** | Community video calls | A professional to check the exercises' answers |
| **4** | Relationship with patients from the main community | Suggestions of materials |
| **5** | Social media groups | Group meeting with a psychologist |
| **6** | Discussion with women after events | Monitor symptoms through medical diaries |
| **7** | Support from co-workers | Physiotherapist's drawings about the exercises |
| **8** | Buddy program | Explanation about medicine and side effects |
| **9** | In-person events | Weekly communication (e.g., via e-mail) |

**Table 5. Overview of elicited social support interventions (i.e., RepGrid elements) ranked by importance**

The findings from our analysis highlight that mentorship is the most significant intervention for peer-to-peer support. In this mentorship model, patients with more experience in managing their condition play a crucial role in guiding those who are newer to the experience. They offer insights and practical advice, helping newcomers navigate the complexities of their health challenges. Additionally, for professional support, the most important intervention identified is the combination of online webinars and Q&A sessions. This format not only grants patients direct access to medical advice but also fosters a learning environment where they can engage in discussions on a broad spectrum of topics related to their condition. Overall, our results emphasize the importance of enabling patients to form strong connections with their peers, facilitating a sense of community and shared understanding. This peer support is complemented by the opportunity for patients to regularly interact with healthcare professionals, which is deemed equally critical. These interactions provide a platform for continuous learning, professional guidance, and reassurance, enhancing the overall management of their condition. The dual approach of integrating both peer and practitioner support through mentorship and educational webinars aligns with the needs of patients, offering them comprehensive support in their health journey.

## Discussion & Conclusion

This study proposed a patient-centered framework that leverages social-technical systems theory and the BPSTM to offer a mixed-method research process to identify design requirements and interventions for health conditions that are affected by social stigma. The framework was theoretically derived and empirically tested with CPP patients.

The theoretical implications of our study are twofold. First, our findings adapt the BPSTM to the context of chronic pelvic pain, with Table 3 mapping out pertinent health determinants for CPP patients and pinpointing three novel determinants—Physiological, Partner Relationships, and Clothing. Our findings provide support for the conceptual premises of viewing health care from a systems-based view (Card 2022). Second, it challenges the existing patient-centered models in literature (Staszak et al. 2021), advocating for a more focused framework that addresses the nuances of social stigma in healthcare. Thus, our study enriches the dialogue on integrating patient-centered frameworks with Sociotechnical Systems theory, highlighting the importance of combined models and interlaced perspectives in tackling complex healthcare challenges

From a managerial perspective, the implications are significant. The framework equips manufacturers with a strategic blueprint to identify design requirements and devise interventions for health conditions that carry a stigma - with many of them being underserved. In a competitive global economy, where innovation and differentiation are key, harnessing opportunities where competition/supply is limited or nonexistent not only makes sense for socio-cultural equity but also positions businesses for robust financial growth and sustainable market leadership. Moreover, our research delineates a set of ten design requirements that are critical for CPP patients and identifies preferred intervention types - priming and social support - for the development of a solution for CPP patients.

Despite the value that our framework offers, it has one main limitation, which is that it considers only Italian patients. Therefore, we note the importance of understanding or considering this when interpreting the findings and their applicability to real-world contexts. Accordingly, the interventions and design requirements we have delineated are intended to assist researchers in crafting their research experiments. In conducting such experiments, we recommend assessing the effects that distinct types of social support interventions may have on health outcomes (e.g., pain interference, sexual well-being, mental well-being). Researchers might, for example, compare the health outcomes of participants who received peer-to-peer support against those who benefited from expert support. An investigation into the relative effectiveness of interventions at different stages of treatment—whether at the beginning, middle, or end—could also yield insightful results. Additionally, analyzing how user engagement acts as a mediator between social support interventions and health outcomes can provide deeper insights into how to design effective mobile health applications.

Our research advances the body of literature by exploring the application of Sociotechnical Systems theory within the domain of mobile health (mHealth) for a patient-centered approach to health-related stigma. By embedding STS theory into mHealth research, we illuminate the dynamics between technological solutions and the social realities of patients. This integration fosters a deeper understanding of how patient-centered care can be enhanced through digital platforms. Furthermore, our study identifies and synthesizes crucial elements of patient-centered design, advocating for their incorporation within a mixed-method design framework. This approach is required for capturing the complex experiences of patients and ensuring that the interventions are responsive to their needs, particularly in the context of stigmatized health conditions. Building upon this, we propose a novel application of the five dimensions of the Biopsychosociotechnical Model, illustrating how they can be effectively intertwined to address the complexities associated with stigmatized health conditions. By considering the biological, psychological, social, technical, and disease dimensions, we present a more nuanced and empathetic approach to health intervention design. In sum, this study paves the way for a transformative paradigm in mHealth, where the convergence of sociotechnical insights and patient-centered care promises to reshape the landscape of healthcare for stigmatized conditions.

*Acknowledgments*

The authors would like to express gratitude to all participants of the study and to Hale UG (haftungsbeschränkt) for their great support.

*Disclosure Statement*